\documentclass[showpacs]{revtex4}

\usepackage{graphicx}
\usepackage{dcolumn}
\usepackage{multirow}
\usepackage{placeins}
\usepackage{amsmath,amssymb,amsfonts,bm}

\begin{document}

\newcommand{\dd}{\text{d}} 

\title{Vibrating systems in Schwarzschild spacetime: towards new experiments in gravitation?}
\author{L.~Bergamin}
\email{bergamin@tph.tuwien.ac.at}
\author{P.~Delva}
\email{pacome.delva@esa.int}
\affiliation{European Space Agency, The Advanced Concepts Team, Keplerlaan 1, 2201 AZ Noordwijk, The Netherlands}
\author{A.~Hees}
\email{aurelien.hees@oma.be}
\affiliation{Observatoire Royal de Belgique (ORB), Avenue Circulaire 3, 1180 Bruxelles, Belgium}

\begin{abstract}
In this paper the effects of vibrations at high frequencies onto a freely falling two-body system in Schwarzschild spacetime are investigated. As reference motion of the same system without vibrations a circular orbit around the central body is considered. The vibrations induce a perturbation on this motion, whose period is close to the orbital period, in agreement with the simpler situation of the Shirokov effect \cite{Shirokov:1973gr}. In general relativity the amplitude of the perturbation is dominated by high velocity effects, which grow linearly in the radius $r$ of the circular orbit, while the leading term surviving the Newtonian limit decays as $1/r$. Thus even for very large radii a significant difference between Newtonian physics and general relativity is found. We give an estimate of this effect for some molecular vibrations of a system orbiting around the Earth.
\end{abstract}

\maketitle


\section{Introduction}

Changing the trajectory of an orbiting spacecraft can be done with internal motions. This is due to the work of the tidal forces experienced by the system. Two-body systems orbiting the Earth have been studied in~\cite{Martinez:1987om,Landis:1991sr,Landis:1992ro}. As most important limitation, large effects in Newtonian physics are resonant effects, where the frequency of change of shape is linked to the motion of the system (e.g.~its orbital frequency or its orientation.). More recently Wisdom~\cite{Wisdom:2003aa} and Gu\'eron et al.\ \cite{Gueron:2005ye,Gueron:2006fq} studied similar situations within general relativity. Within this more general setup non-resonant effects can become large. Thus, relevant deviations from the free motion can be obtained for frequencies of the change of shape completely independent of the orbital motion.

In this paper we investigate a vibrating two body system similar to the one of Ref.~\cite{Gueron:2006fq}. As reference motion a circular orbit is considered (as opposed to a radial free fall as in Ref.~\cite{Gueron:2006fq}) and it is shown that this leads to a systematic deviation between the Newtonian result and general relativity even for large radii of the orbit (i.e.\ weak gravitation) since the leading perturbation in general relativity is a high-velocity effect, where the vibrational velocity acts as the key parameter. In particular, the maximal deviation per orbit is found to \emph{increase} with increasing radius of the orbit. Rather than with the result of Ref.~\cite{Gueron:2006fq} our result is related to the so-called Shirokov effect \cite{Shirokov:1973gr}. Still, the latter work just considers one perturbation at a single time instance and not a continuous perturbation as done in this work. Accordingly we find effects which are at least three orders of magnitude bigger than the effect reported in \cite{Shirokov:1973gr}.

The paper is organized as follows: the following section presents the model and the geometry used throughout. In Section~\ref{dcs} a simplified model is studied in detail and it is shown that in general relativity the total deviation from the free motion increases linearly with the distance from the central body whereas the dominant term surviving in the Newtonian limit decreases. In the subsequent section, we make a numerical study of the full system and we show that the effect found for the simplified case still dominates the behavior. In Section~\ref{secExp} we comment possible experiments and finally our conclusions are presented in Section~\ref{conclusions}.


\section{The model} \label{s:model}

\begin{figure}[t]
\begin{minipage}{0.48\linewidth}
\begin{center}
		\includegraphics[width=0.8\linewidth]{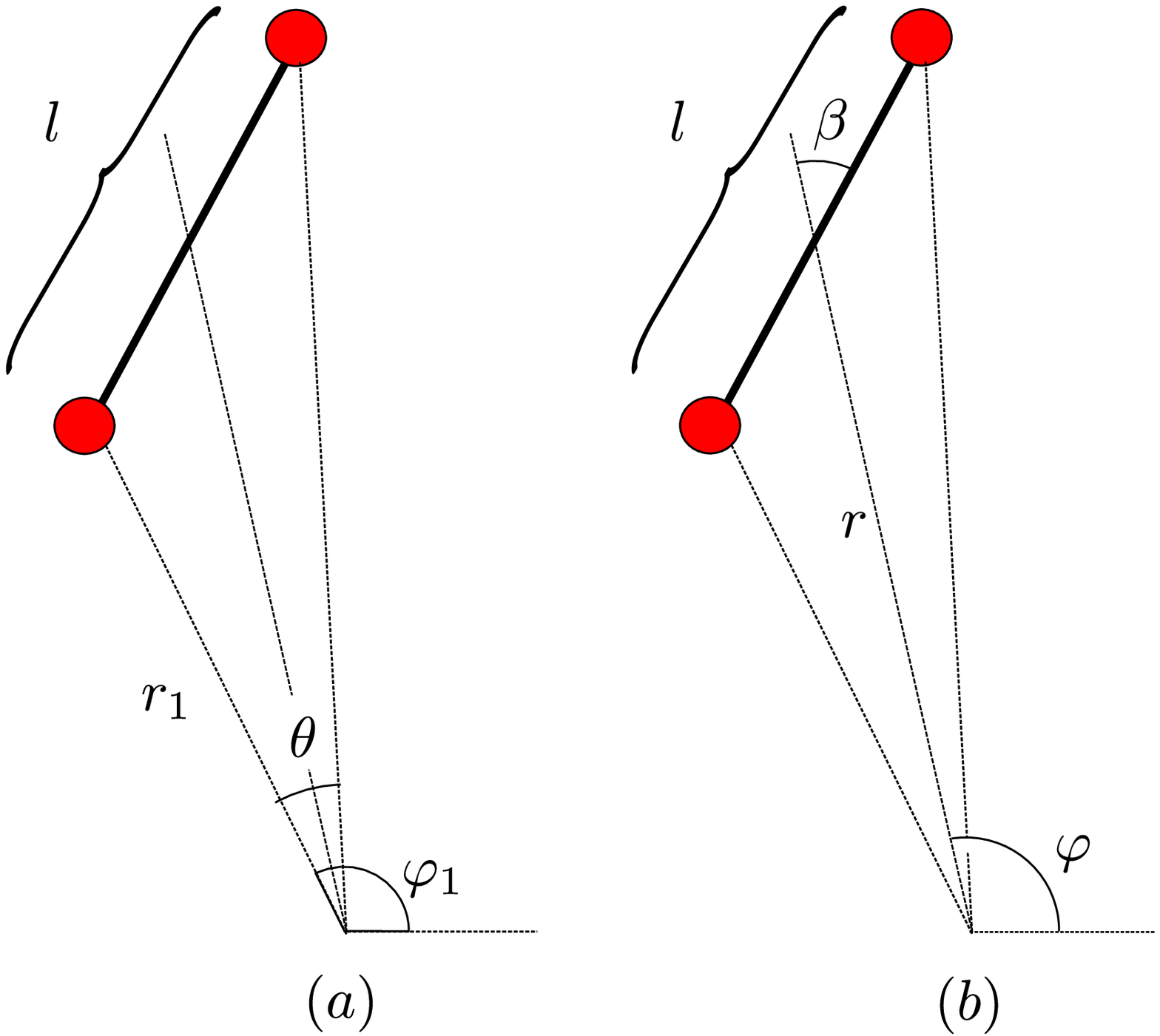}
	\caption{Representation of our model and the different variables used.}
	\label{figSys1}
\end{center}
\end{minipage}
\begin{minipage}{0.48\linewidth}
\begin{center}
		\includegraphics[width=0.8\linewidth]{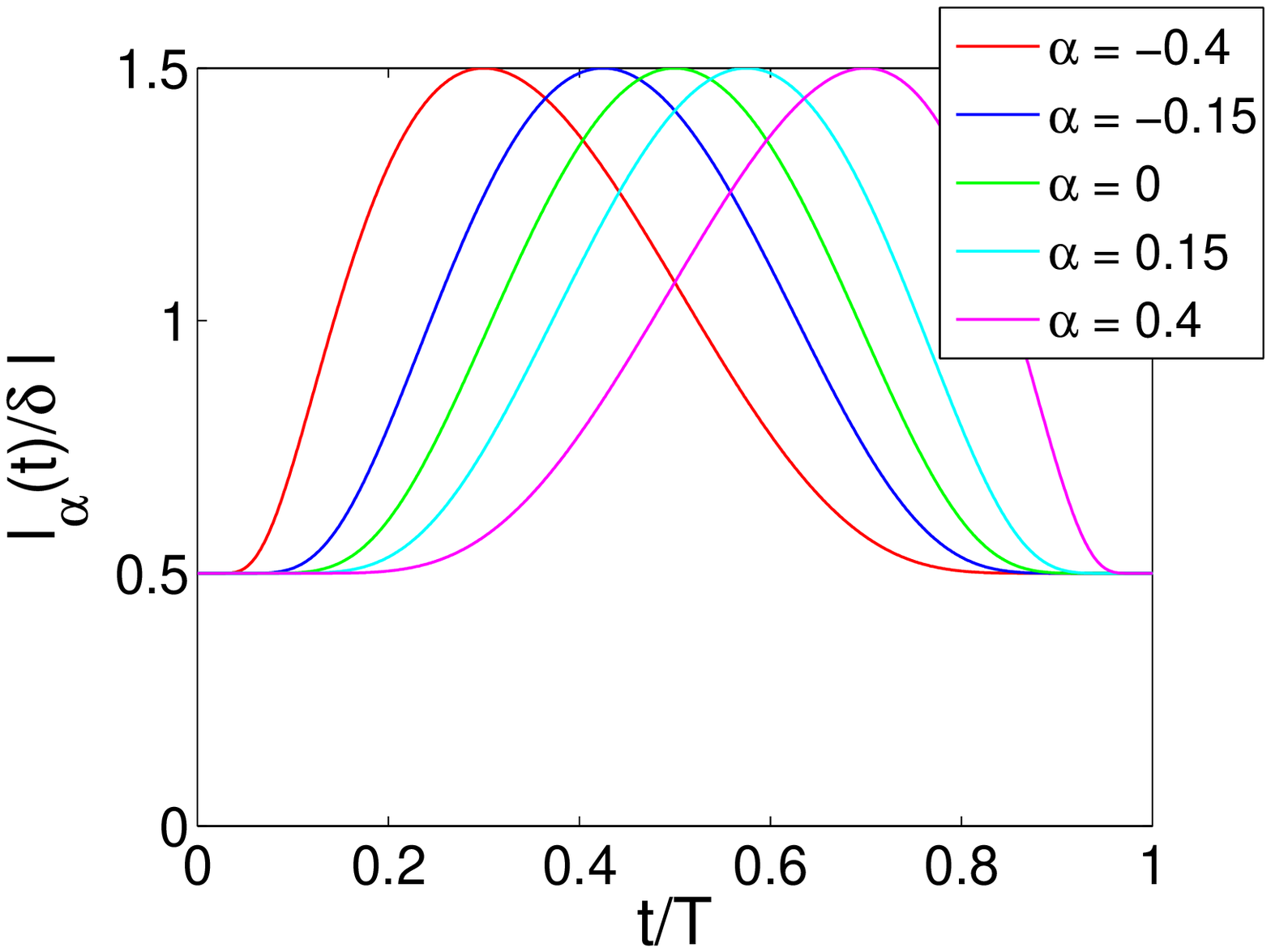}
	\caption{Profile of the constraint.}
	\label{ProfCons}
\end{center}
\end{minipage}
\end{figure}

Following the previous works~\cite{Wisdom:2003aa,Gueron:2005ye,Gueron:2006fq}, a vibrating or oscillating system is implemented as a collection of point masses whose relative positions are related by time dependent constraints. The specific model used here essentially is equivalent to the one of Ref.~\cite{Gueron:2006fq}: a two-body system made from two test masses connected by a massless tether, whose length $l(t)$ is imposed by an oscillating constraint. To further simplify the situation the two masses are always considered to be equal and put to unity in all calculations. This system shall orbit around a central body, which is described by means of Schwarzschild geometry. We assume that the two masses have the same orbital plane, where the oscillations take place; therefore in Schwarzschild coordinates, the motion of each of the two point masses can be defined in terms of the radial and azimuthal coordinates $(r_i,\varphi_i)$ (where $i=1,2$), while the polar angle angle is dropped. The system may be fully described in terms of the variables $(r_1,\varphi_1)$, the relative angle $\theta = \varphi_2-\varphi_1$ and the constraint $l(t)$ (see Fig.~\ref{figSys1}(a)). However, in many situations it is useful to use instead the coordinates of the geometrical center of mass $(r,\varphi)$, the relative angle between {\boldmath$r$ and $l$}, $\beta$, and the constraint $l(t)$. This situation is depicted in Fig.~\ref{figSys1}(b).

In Ref.~\cite{Gueron:2006fq} it has been shown for a radial free fall that asymmetrical oscillations induce a quite different change in the motion of this system than symmetrical do. To assess the same question for closed orbits, a similar profile of the constraint including a potential asymmetry is used here; however, it will be shown below that the exact profile of the constraint is unimportant. Our constraint is described by four parameters: its frequency  $\tilde{\omega}=1/\tilde{T}$, its amplitude $\delta_l$, its minimum length $l_0$ and an asymmetry parameter $\alpha$. The asymmetry parameter $\alpha$, taking values in the range $[-1,1]$, indicates how much the constraint fails to be symmetric with $\alpha = 0$ being the symmetric case\footnote{The expression of the constraint used here is based on the one used in Ref.~\cite{Gueron:2006fq}: $l(t;\tilde{\omega},\alpha,\delta_l,l_0)=l_0+\delta_l \exp\left[\frac{(1+\alpha-2\tilde{\omega} t)^2 }{(1+\alpha^2)\tilde{\omega} t(-1+\tilde{\omega} t)}\right]$ for $t\in[0,\tilde{T}]$ and periodic with period $\tilde T$.}. Unlike Ref.~\cite{Gueron:2006fq}, we need to introduce a minimal length $l_0$ between the two masses in order to avoid divergent angular momentum of the spinning two-body system. The profile of the constraint is shown in Figure~\ref{ProfCons}.

The constraint $l$ relates the coordinates of the two masses as
\begin{equation}
l^2=r_1^2+r_2^2-2r_1r_2\cos\theta\ .
\end{equation}
Since $l/r$ and consequently also $\theta$ is always kept small, $r_2$ in very good approximation can be expressed as
\begin{equation}
r_2=r_1 + \sqrt{l^2(t)-r_1^2\theta^2}=r_1+\delta r_{12}\ .
\end{equation}

In many explicit calculations the above system is yet too complicated. To reduce clutter a simplified situation will be considered. In this system an additional constraint is imposed which enforces $\varphi_1=\varphi_2=\varphi$ and thus $\theta=0$. We call this system the double constrained system (by opposition to the full system).

We point out that no corrections from spacetime curvature have been taken into account in the calculations of the relative coordinates between the two masses. Though our ansatz mainly will be justified by the final result, we would like to comment about these issues at this place already. Most importantly it should be realized that we intend to place the constrained system exclusively in regions of weak gravity, where $r_s/r \ll 1$, and thus the ensuing errors are expected to be small. Furthermore it will be shown in Section~\ref{sec:analytical} by means of the expansion of the equation of motion of our vibrating system that a systematic difference exist between general relativity and the Newtonian theory, which is not affected at all by this simplification. Thus, this small error just affects the numerical results presented in the paper; our main conclusion based on analytical expansions remains unchanged.

Still, the interpretation of the implementation of the constraint and the resulting deviation $\Delta r$ from the reference motion in terms of the Schwarzschild coordinate $r$ may be questioned. However, we are not interested here on finding an experimental protocol to implement a particular constraint, or to describe a realistic constraint. If so, one could implement the constraint in the Fermi-Walker coordinates~\cite{Misner:1962} of the reference motion, which in good approximation are comoving coordinates of the vibrating system. This would not change our results qualitatively. Moreover, the deviation $\Delta r$ is a coordinate distance between two defined events along the perturbed and the non-perturbed trajectories. We will express it in term of radar distance, which is then independent of the particular coordinates used in the calculations.

Let us now write down explicitly the relevant actions of our system in general relativity. As explained above, we assume the spacetime geometry to be described by the Schwarzschild metric, and we implement the constraint in Schwarzschild coordinates. Thus the action becomes\footnote{Geometrical units ($c=G=1$) are used in this paper.}:
\begin{equation}\label{action}
S=-\int \dd t  \left[ \sqrt{L_1} + \sqrt{L_2} +\lambda \left(r_2-r_1-\delta r_{12}\right) \right] ,
\end{equation}
where $\lambda$ is a Lagrange multiplier, $t$ is the Schwarzschild time and
\begin{equation}
L_i=\left(1-\frac{r_s}{r_i}\right)-\left(1-\frac{r_s}{r_i}\right)^{-1}\dot{r}_i^2-r_i^2\dot{\varphi}_i^2 ,
\end{equation}
with $r_s$ being the Schwarzschild radius of the central mass, and $\dot{()} \equiv \dd / \dd t$.

In the following sections, we study the effect of the oscillations of the length of the constraint, $l(t)$, onto the motion of the two-body system. We compare the trajectory of the vibrating system with a reference motion, which is the motion of the same two-body system with identical initial conditions but without vibration. For the reference motion the simplest situation, namely circular orbits, are chosen and the two-body system always is considered to be aligned in the radial direction. We emphasize that the reference motion is not a geodesic. However, it is straightforward to show that $\dot{\phi}$ is constant for the reference motion. The initial conditions of the non-vibrating and the vibrating systems are denoted by: $r(0)=r_0$, $\varphi(0)=0$, $\theta(0)=0$, $\dot{r}(0)=0$, $\dot{\varphi}(0)=\omega$ and $\dot{\theta}(0)=0$. By imposing the reference motion to be circular we get from the conservation of angular momentum a relation between $r_0$ and $\omega$; for the following the explicit relation between $r_0$ and $\omega$ is not important so we do not reproduce it here.


\section{The double-constrained system}
\label{dcs}
We first study the double constrained system, which is described by the action~(\ref{action}) with the additional constraint $\theta=0$. It turns out that this system is much simpler than the full system and even allows to make analytical expansions, since
\begin{itemize}
 \item it contains one degree of freedom less than the full system;
 \item it allows a much simpler implementation of the constraint $l(t)$;
 \item it includes only two frequencies (the frequency of the constraint $l(t)$ and the orbital frequency) instead of three in the full system, which in addition includes the frequency of the oscillation of the two-body system around its geometrical center of mass.
\end{itemize}
We hasten to add that the motivation to study the double-constrained system is of purely theoretical nature, mainly the fact that this system can be treated analytically. We do not claim at this point that this specific system easily can be implemented in a real experiment.
Nonetheless, it will be shown in Section \ref{sec:full} that the differences between the double-constrained and the full system remain very small. This justifies to study the double-constrained system more in detail since the dynamics of the full system are clearly dominated by the same effects as the ones to be worked out in this section.


\subsection{Linearization of the equations of motion}
Of course, the exact equations of motion can be integrated numerically, which however can be quite time consuming, especially for large radii. Without use of extensive computer resources it is not possible to integrate the equations for a system orbiting the Earth. Fortunately, the double constrained system can be treated analytically in very good approximation. To do that, the equations of motion are linearized around the reference motion, which we have defined to be the motion of a non-vibrating system with the same initial conditions. As pointed out above, this reference motion is assumed to be a circular orbit characterized by the radius $r_0$ and the orbital frequency $\omega$. The parameters of the vibrating system become $r=r_0+\delta r$ and $\varphi=\omega t+\delta \varphi$ and yield equations in the deviation variables $\delta r$ and $\delta \varphi$, which are then expanded to first order. One of the two resulting equations is of the type

\begin{equation} \label{e:lin}
A(t)\delta \ddot{r}+B(t)\delta \ddot{\varphi}=C_1(t)\delta \dot{r} + D_1(t) \delta r +E_1(t) \delta  \dot{\varphi}  + F_1(t)\ .
\end{equation}

Since the angular momentum
\begin{equation}
L=\dot{\varphi}\left(\frac{r_1^2}{\sqrt{L_1}}+\frac{r_2^2}{\sqrt{L_2}}\right)
\end{equation}
is conserved, the linearized angular momentum allows to substitute $\delta \dot{\varphi}$ in~(\ref{e:lin}), which yields
\begin{equation}\label{linear}
A(t)\delta \ddot{r}+B(t)\delta \ddot{\varphi}=C(t)\delta \dot{r} - D(t) \delta r  - F(t)\ .
\end{equation}
Here $A(t)$, $B(t)$, $C(t)$, $D(t)$ and $F(t)$ are complicated functions but they are all periodic with the period of the constraint $\tilde{T}=1/\tilde{\omega}$. At this point it is important that our system only includes two frequencies, $\tilde \omega$ and the orbital frequency $\omega$. As is confirmed by integrating the exact equations, $\delta r$ oscillates with a period of the order of $\omega$. Since $\omega<<\tilde{\omega}$, $\delta r$ does not relevantly change during one period $\tilde T$ and thus all coefficients $A(t)$, $B(t)$, $C(t)$, $D(t)$ and $F(t)$ may be averaged over one period of the constraint, i.e.\ $A(t)$ is replaced by $\bar{A}=\frac{1}{\tilde{T}} \int_0^{\tilde{T}} A(t)dt$ etc. Using this simplification, one can show that $\bar{B}=\bar{C}=0$, and the solution of the equation of motion simply becomes
\begin{equation} \label{omegaprime}
\delta r=\frac{\bar{F}}{\bar{D}}\left(\cos \omega't-1\right)\ ,
\end{equation}
where $\omega'^2=\bar{D} / \bar{A}$. A numerical integration shows that $\bar{D} / \bar{A}$ is positive, and that $\omega' \sim \omega$. Moreover, $|\delta r| \ll r_0$ implies that $\bar F/(\bar D r_0) \ll 1$. Then, the resulting trajectory of the vibrating system is an ellipse of eccentricity $e \approx \bar F/(\bar D r_0)$. The advantage of this formulation is that we only need to perform an integration of the three functions $A$, $D$ and $F$ over one period of oscillation instead of integrating the full equation~\eqref{linear} over the desired time of evolution, e.g.\ one revolution around the central body. In this way we directly obtain an expression for the maximal value $\Delta r$ of the deviation from the reference motion:
\begin{equation}\label{drLin}
\Delta r=\frac{2\bar{F}}{\bar{D}}\ .
\end{equation}
The corresponding radar distance $\Delta l$ can be obtained by integrating the relation $\dd l^2 = (1-rs/r)^{-1} \dd r^2$ between $r_0$ and $r_0+\Delta r$. The radar distance does not depend on the particular coordinate system we used for the calculation, and \emph{in principle} can be determined by instruments on-board the spacecraft.


\subsection{Analytical calculation}
\label{sec:analytical}
\begin{figure}[t]
	\centering
		\includegraphics[width=0.5\textwidth]{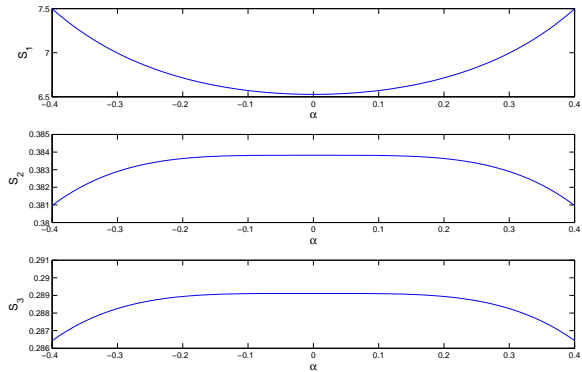}
	\caption{Numerical evaluation of $S_1(\alpha)$, $S_2(\alpha)$ and $S_3(\alpha)$.}
	\label{figS}
\end{figure}
The result \eqref{drLin} still does not allow a full analytic treatment, since the functions $F$ and $D$ cannot be integrated analytically. 
Still, further expansions of the equation of motion \eqref{linear} are possible since the two parameters $l(t)/r_0$ and $\dot l/c$ both remain small~\footnote{Since an expansion in velocities is performed, the speed of light $c$ is explicitly written in this section to allow a simple identification of the orders of expansion.}. We do not expand in the parameter $r_s/r_0$ as this is not necessary to obtain our result. Notice however that all our discussions take place in a regime of weak gravity, where $r_s/r_0 \ll 1$.

In our calculations, $\dot l/c$ typically is of the order of $10^{-4}$--$10^{-5}$, while $l/r_0$ ranges from $10^{-5}$ at relatively small radii up to $10^{-12}$ in Earth orbit (cf.\ also the numbers presented in Sect.~\ref{secExp}). Then we obtain:
\begin{eqnarray}\label{analExp}
\dfrac{\Delta  r}{r_0} & = & \frac{\delta_l^2\tilde{\omega}^2}{c^2}\frac{S_1(\alpha)}{\left(1-\frac{r_s}{r_0}\right)\left(1-\frac{3r_s}{r_0}\right)} \nonumber\\
&+& \dfrac{5}{2} \dfrac{\delta_l}{r_0} \frac{1-\frac{9r_s}{r_0}}{1-\frac{3r_s}{r_0}} \left( 2 \dfrac{l_0}{r_0} S_2(\alpha) + \dfrac{\delta_l}{r_0} S_3(\alpha) \right)  \\
&+&\mathcal O \left(\frac{\dot{l}^2}{c^2}\frac{l^2}{r_0^2} \right)+\mathcal O \left(\frac{\dot{l}^4}{c^4} \right) +\mathcal O \left(\frac{l^4}{r_0^4} \right) \ . \nonumber
\end{eqnarray}
$S_1$, $S_2$ and $S_3$ are functions that depend on the asymmetry parameter and are represented in Fig.~\ref{figS}. The deviation $\Delta r$ contains two kinds of terms: a high velocity effect on the first line and a purely gravitational effect on the second line. These two effects behave differently with $r_0$. In the following, we will assume that $l/r_s \lesssim 10^{-3}$ and $r_s/r_0 \ll 0.01-0.1$~\footnote{In this situation one could consider alternatively to Eq.~\eqref{analExp} an expansion with respect to $l/r_S$ instead of $l/r_0$, which would lead to the same conclusions. We have chosen the form of Eq.~\eqref{analExp} since this expansion also covers the case $r_0\gg r_S \approx l$.}; then the high velocity effect is much bigger than the purely gravitational effect. This dominance increases with increasing radius, since the velocity effect grows linearly with $r_0$~\footnote{This is possible since the integration time increases as $r_0^{3/2}$. Per unit time, the relativistic effect decays like $1/\sqrt{r_0}$, while the purely gravitational effect scales like $1/r_0^{5/2}$.}, while the purely gravitational effect decays like $1/r_0$. Keeping only the leading term in~\eqref{analExp} while expanding in $r_s/r_0$, we obtain:
\begin{equation} \label{e:dev}
\dfrac{\Delta r}{r_0} \simeq \frac{\delta_l^2\tilde{\omega}^2}{c^2} S_1(\alpha) .
\end{equation}
The corresponding radar distance, to zeroth order in $r_s/r_0$, is $\Delta l \simeq \Delta r$. Then, to this order the deviation~\eqref{e:dev} does not depend on the choice of the coordinate system we used for the calculation. Also it is seen that the deviation is determined solely in terms of the change of the length of the constraint, $\delta_l$, and its frequency; as mentioned already above the exact form of the constraint remains unimportant.

The Newtonian limit of equation~\eqref{analExp} is obtained by taking the limit $\dot{l}/c \rightarrow 0$ and $r_s/r_0 \rightarrow 0$:
\begin{equation}\label{drNew}
\dfrac{\Delta r_{N}}{r_0} = \dfrac{5}{2} \dfrac{\delta_l}{r_0} \left( 2 \dfrac{l_0}{r_0} S_2(\alpha) + \dfrac{\delta_l}{r_0} S_3(\alpha) \right) \ .
\end{equation}
Given the fact that the leading effect in general relativity is a high velocity effect it is evident that the differences between general relativity and Newtonian gravity are much more fundamental than their mere difference in the interpretation of the radial coordinate (we point out again that all the results apply in regions with weak gravitation and small velocities of the system as a whole, where this difference is small). Since the effect due to the velocity term grows with increasing radius, while the difference in the interpretation of the radial coordinate decreases, this conclusion applies to any desired accuracy if only gravity is weak enough.

Before studying $\Delta r$ in detail, we want to show that the resulting motion of the vibrating system to a very good approximation is an ellipse. From~\eqref{omegaprime} the frequency or period of the perturbation $\delta r$ can be calculated. With $T$ being the time of revolution of the reference motion, the period of the perturbation is found as
\begin{equation}
\begin{split}
T'&=\frac{T}{\sqrt{1-\frac{3r_s}{r_0}}}\left(1 + k_1\frac{\delta_l^2\tilde{\omega}^2}{c^2}+ \mathcal O \left(\frac{l^2}{r_0^2}\right) +\mathcal O \left(\frac{\dot{l}^4}{c^4} \right) \right)\\ &\simeq T\left(1+\frac{3r_s}{2r_0}+k_1\frac{\delta_l^2\tilde{\omega}^2}{c^2}\right)\ ,
\end{split}
\end{equation}
where $k_1$ is a function of the ratio $r_s/r_0$. Note that $k_1 \rightarrow 1$ when $r_s/r_0 \rightarrow 0$, so the perturbation in the orbital period stays even in a weak gravity field. To the same order in the expansion, the advance of the periapsis is given by
\begin{equation}
\begin{split}
\Delta \varphi &\simeq\frac{2\pi}{\sqrt{1-\frac{3r_s}{r_0}}}\left(1+k_2\frac{\delta_l^2\tilde{\omega}^2}{c^2}\right)-2\pi \\ &\simeq 2\pi \left(\frac{3r_s}{2r_0}+k_2\frac{\delta_l^2\tilde{\omega}^2}{c^2}\right)\ ,
\end{split}
\end{equation}
where $k_2$ is again a function of $r_s/r_0$, which for large radii is of order 1. At large radius, the second term dominates over the first one but in absolute terms stays very small (cf.\ Sect.~\ref{secExp} for further discussions). Thus the resulting motion indeed is an ellipse with a small advance of the periapsis.

It is interesting to look at the conservation of energy. In general relativity the Hamiltonian $H=H_1+H_2$, with
\begin{equation}
H_i=\frac{1-\frac{r_s}{r_i}}{\sqrt{L_i}} ,
\end{equation}
may be used as a definition of ``energy''. Obviously, this Hamiltonian is conserved for the non-vibrating system. Together with oscillations, the energy changes during each cycle of the oscillation. Nonetheless, looking only at the starting or end points of the vibrations, $t = (0, \tilde T, 2\tilde T, \ldots)$, the energy remains conserved to second order in $l/r_0$ and to second order in $\dot{l}/c$. This means that the variation of energy (if there is any at all) is of higher order and therefore remains very small. This analysis is confirmed by numerical simulations; however, due to numerical errors, it is hard to quantify its actual change.

In conclusion the vibrations do not change the energy of the system considerably. Thus, although the orbit of the vibrating system geometrically is an ellipse, the motion of the system is different from the geodesic motion described by the same ellipse, since the energies (and thus the velocities) of the two situations are different.

In the remainder of this section we comment on the relation of our system to the so-called Shirokov effect \cite{Shirokov:1973gr, Melkumova:1990rp, Vladimirov:1981rp}. In the Shirokov experiment, a non-vibrating and point-like body is perturbed from a reference circular orbit. Then its trajectory oscillates around its reference orbit, with a different period if the movement is along the radial or the zenithal directions. The experiment consist to find the difference between these two periods. For a radial perturbation, Vladimorov has shown that the oscillation is due to the quasielliptical nature of the orbit, with a general relativistic Mercury type displacement of the periapsis~\cite{Vladimirov:1981rp}. The period and the advance of the periapsis of this orbit can be compared to the one we find for the vibrating system, except that for the latter there are supplementary corrections due to the oscillation velocity. However, these two orbits are located differently compared to their circular reference orbit: the orbit of the non-vibrating body in the Shirokov experiment intersects with reference orbit twice per orbital period, whereas the orbit of the vibrating system crosses its reference orbit only once per orbital period. This is due to the small correction $- \bar{F} / \bar{D}$ contained in Equation~\eqref{omegaprime}.

\begin{figure}[t]
	\centering
	\includegraphics[width=0.8\textwidth]{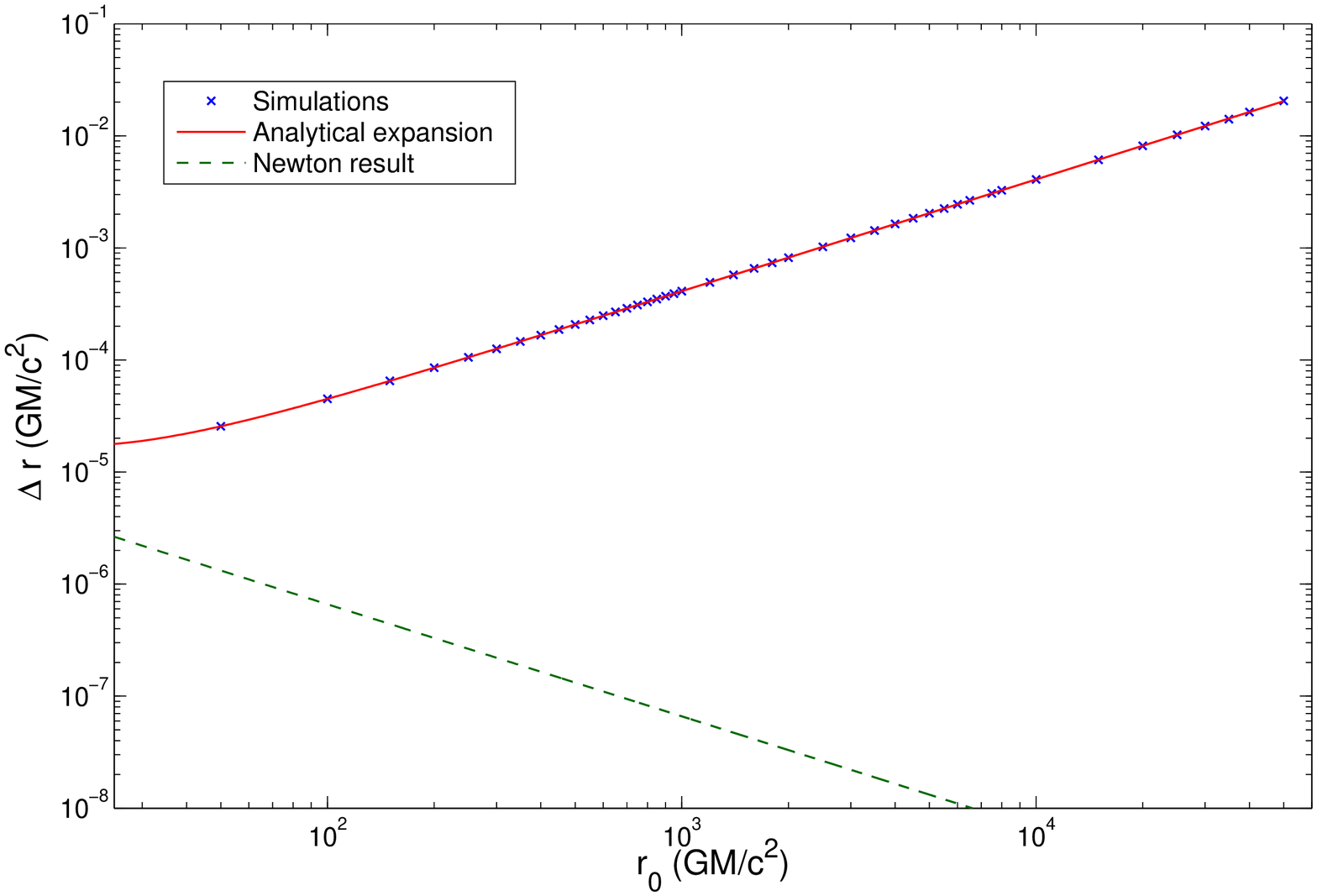}
	\caption{$\Delta r$ as a function of $r_0$ for $\tilde{\omega}=0.05$, $\alpha=0$, $\delta_l=5 \times 10^{-3}$ and $l_0=5 \times 10^{-3}$: comparison of the numerical simulations and the analytical expansion (\ref{analExp}.) In addition the leading purely gravitation term has been included in the figure to illustrate the difference between general relativity and Newtonian gravity.}
	\label{figDrR0}
\end{figure}

Moreover, in~\cite{Shirokov:1973gr}, Shirokov considers as given the maximum deviation of a trajectory perturbed in the radial direction of a non-vibrating body from its reference circular orbit; this is very different from the situation we describe where the maximum deviation is the quantity we want to determine. In fact, in the Shirokov experiment, the only thing that differs between the reference and the perturbed body are their initial conditions, whereas in our situation the initial conditions of the non-vibrating and the vibrating bodies are the same; the difference is in their internal movement. Then Shirokov calculates the coordinate distance $\xi$ between the body perturbed in the radial direction and a body perturbed in the zenithal direction after $n$ orbital periods. For $n=10$ orbits he obtains $\xi \simeq 10^{-6}$~cm. This is at least 3 orders of magnitude below the effect we find for vibrating molecules after one orbit (see section~V).


\subsection{Comparison and results}

In this section results from the two presented approaches, the numerical integration of the complete equations of motion and the analytical result from a systematic expansion \eqref{analExp}, shall be compared.

Figure~\ref{figDrR0} presents\footnote{We work with adimensionnal quantities which means that the lengths are expressed in units of $GM/c^2$, the frequencies are expressed in units of $c^3/GM$, the time is expressed in unit of $GM/c^3$ and the velocities are expressed in units of $c$.} in a log-log scale the numerical analysis of the full equations and the analytical result for $\Delta r$ as a function of $r_0$. For comparison the Newtonian result is included as well. As can be seen, the general relativistic result is perfectly linear above $r_0 \approx 100$; below that value strong gravity effects appear due to the corrections in $r_s/r_0$ in the relation (\ref{analExp}). Furthermore it is seen that the analytical result fits perfectly well the numerical data. Considering the Newtonian result the curve is linear in a log-log scale which fits the $1/r_0$ behavior as found in Eq.~(\ref{drNew}). We can clearly see on this figure the divergence between the classical approach and the relativistic one.

\begin{figure}[t]
\begin{center}
		\includegraphics[width=0.8\linewidth]{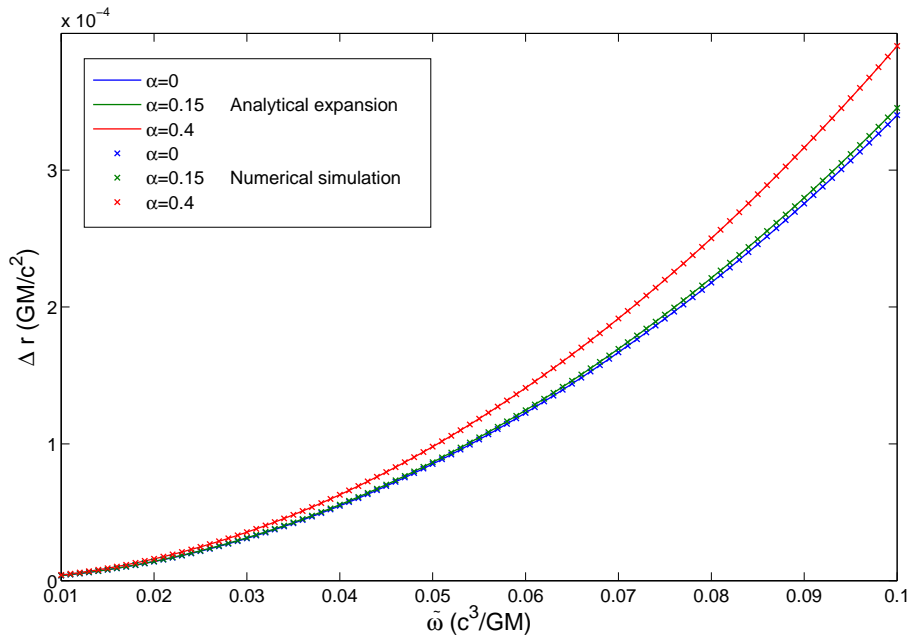}
	\caption{Comparison of the numerical simulations with the analytical result~: $\Delta r$ as a function of $\tilde{\omega}$ for different values of $\alpha=\{ 0,0.15,0.4 \}$, $r_0=200$, $\delta_l=5 \times 10^{-3}$ and $l_0=5 \times 10^{-3}$. The results for negative $\alpha$ are exactly the same as those for positive values.}
	\label{figDrOmega}
\end{center}
\end{figure}

In Figure \ref{figDrOmega} the variations of the frequency of the constraint, $\tilde{\omega}$, and the asymmetry parameter, $\alpha$, for fixed radius $r_0$ are presented. Again the analytical result is compared with the full numerical integration and the quadratic behavior in the frequency is confirmed. As can be inferred from the evaluation of $S_1(\alpha)$ in Fig.~\ref{figS}, $\Delta r$ increases if the asymmetry parameter is chosen different from zero. However, the figure clearly displays that an increase in the frequency is preferable to increase the effect. It is important to point out the differences of this result compared to a radial fall of the same system \cite{Gueron:2006fq}: there, the asymmetry parameter plays a central role since no effect is found at $\alpha = 0$, furthermore the result in that situation is asymmetric for  $\alpha \rightarrow - \alpha$ while it is symmetric here. In conclusion one can say that the potential for asymmetric constraints in circular orbits is very limited. We also mention that the authors of Ref.~\cite{Gueron:2006fq} found a characteristic plateau when varying the frequency. It should be pointed out that this ``plateau'' in the current context should rather be seen as a linear behavior since here we integrate the effect over a certain fixed time rather than a fixed number of oscillations as done in \cite{Gueron:2006fq}. Still, the circular motion shows a stronger dependence on the frequency, since it increases like $\tilde \omega^2$ for increasing frequency. Indeed, the effect found in Ref.~\cite{Gueron:2006fq} is of order $\mathcal O(\dot l \dot r_0/c^2) \propto \tilde \omega$, where $\dot r_0$ is the velocity of the radial free fall of the non-vibrating system, to be compared with the effect from the circular case, of order $\mathcal O(\dot{l}^2/c^2) \propto \tilde \omega^2$.

\begin{figure}[t]
\begin{minipage}{0.48\linewidth}
\begin{center}
		\includegraphics[width=1.0\linewidth]{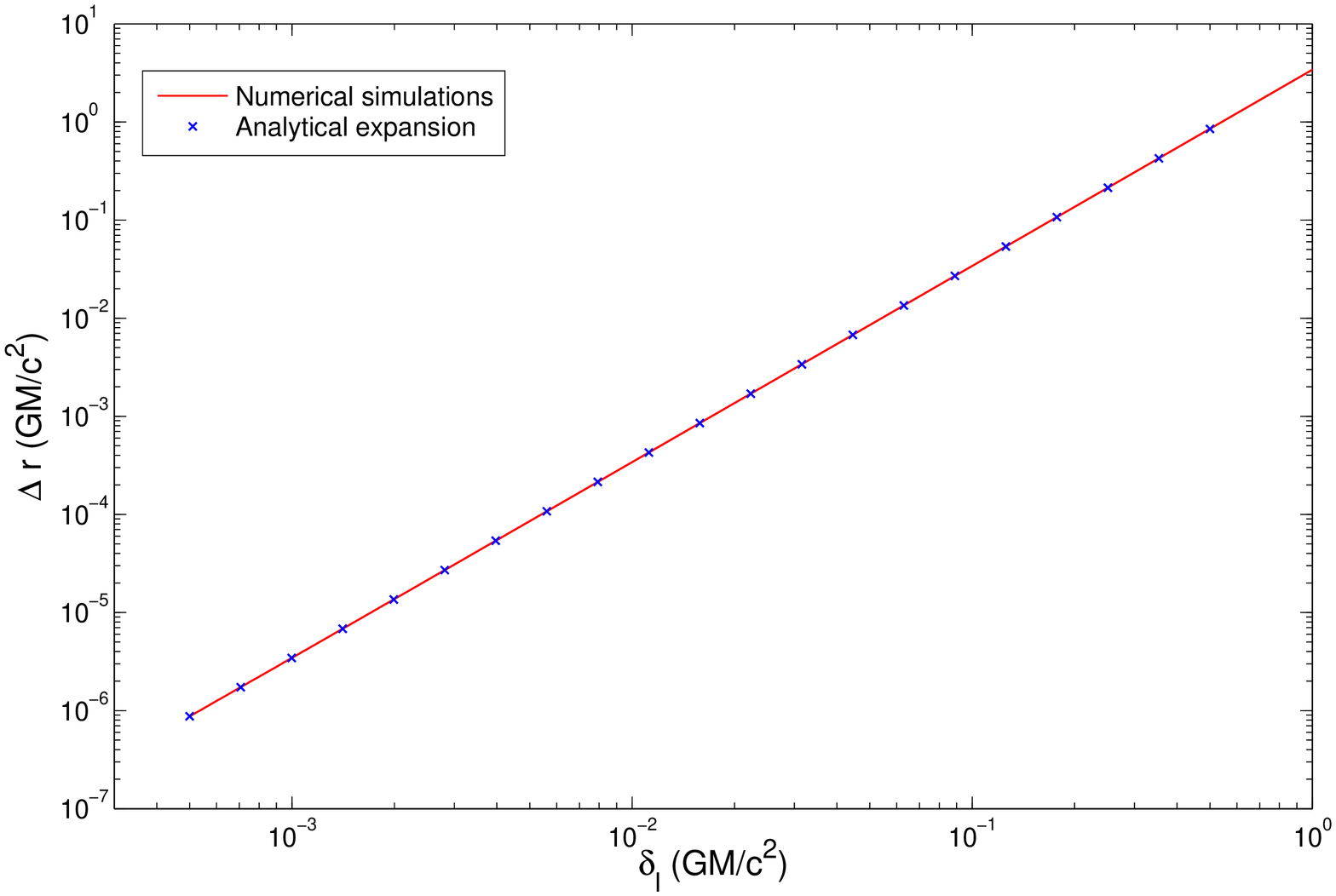}
\end{center}
\end{minipage}
\begin{minipage}{0.48\linewidth}
\begin{center}
		\includegraphics[width=1.0\linewidth]{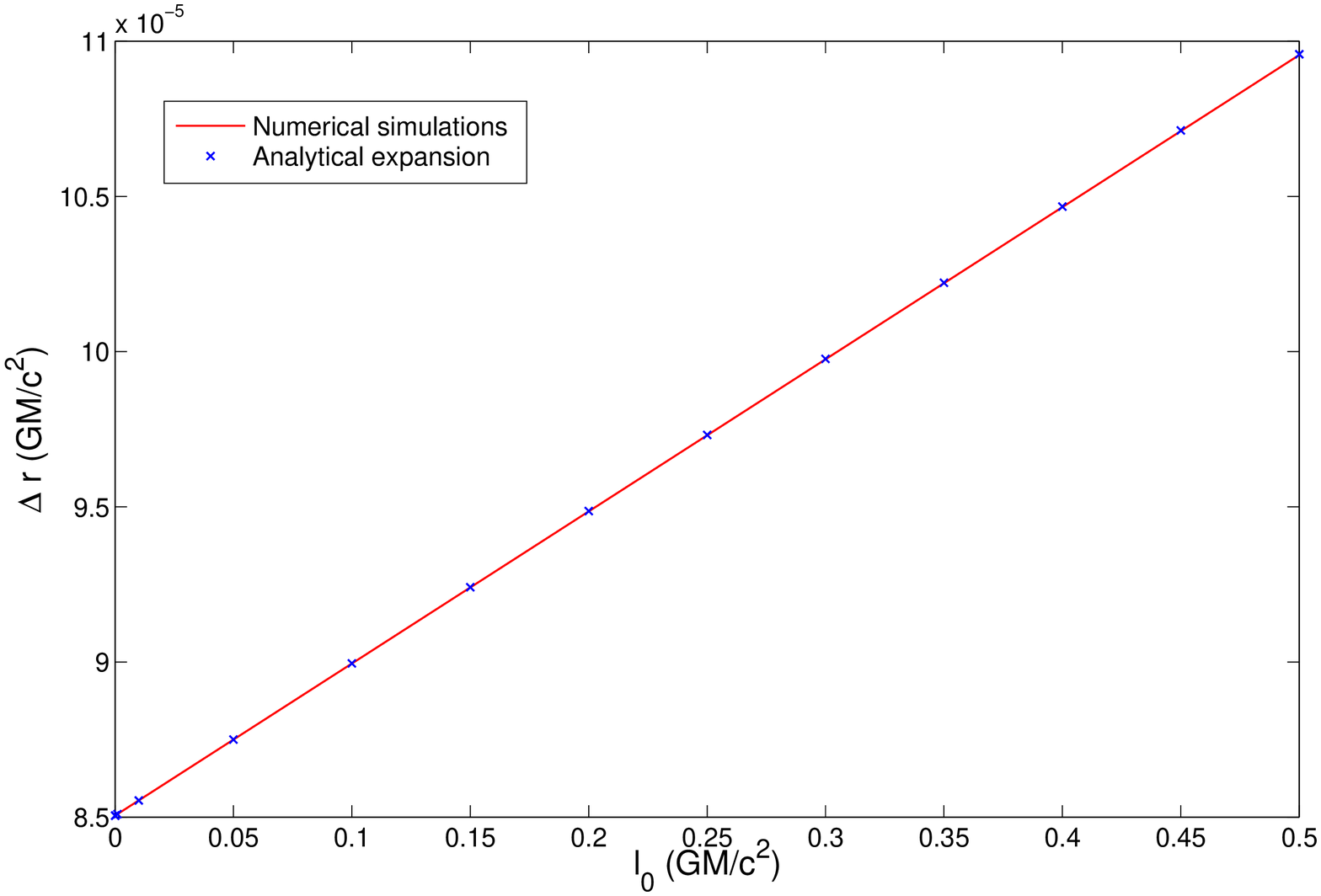}
\end{center}
\end{minipage}
	\label{figDrDl}
	\caption{Comparison of the numerical simulations with the analytical expansion~(\ref{analExp})~: (left) $\Delta r$ as a function of $\delta_l$ for $r_0=200$, $\tilde{\omega}=0.05$, $\alpha=0$ and $l_0=5 \times 10^{-3}$; (right) $\Delta r$ as a function of $l_0$ for $r_0=200$, $\tilde{\omega}=0.05$, $\alpha=0$ and $\delta_l=5 \times 10^{-3}$.}
\end{figure}

Also, we can compare the results from varying the amplitude of oscillations~$\delta_l$. Figure \ref{figDrDl} shows $\Delta r$ as a function of $\delta_l$ obtained from numerical simulations and from the analytical expansion. Again the analytical result fits perfectly well the numerical simulation, confirming the quadratic behavior in this variable

Finally, the behavior of $\Delta r$ for different values of $l_0$ is presented. Figure \ref{figDrDl} reproduces the linear behavior in this variable with a good agreement of numerical data and analytical expansion. Comparing the Figures \ref{figDrDl} and \ref{figDrDl} it is seen that $l_0$ has a small influence on the result in comparison to $\delta_l$. Thus in practice it is important that the system vibrates with a large amplitude, while the minimal distance of the two point masses is of minor relevance.


\section{The full system}
\label{sec:full}

\begin{figure}[b]
\begin{minipage}{0.48\linewidth}
\begin{center}
		\includegraphics[width=1.0\linewidth]{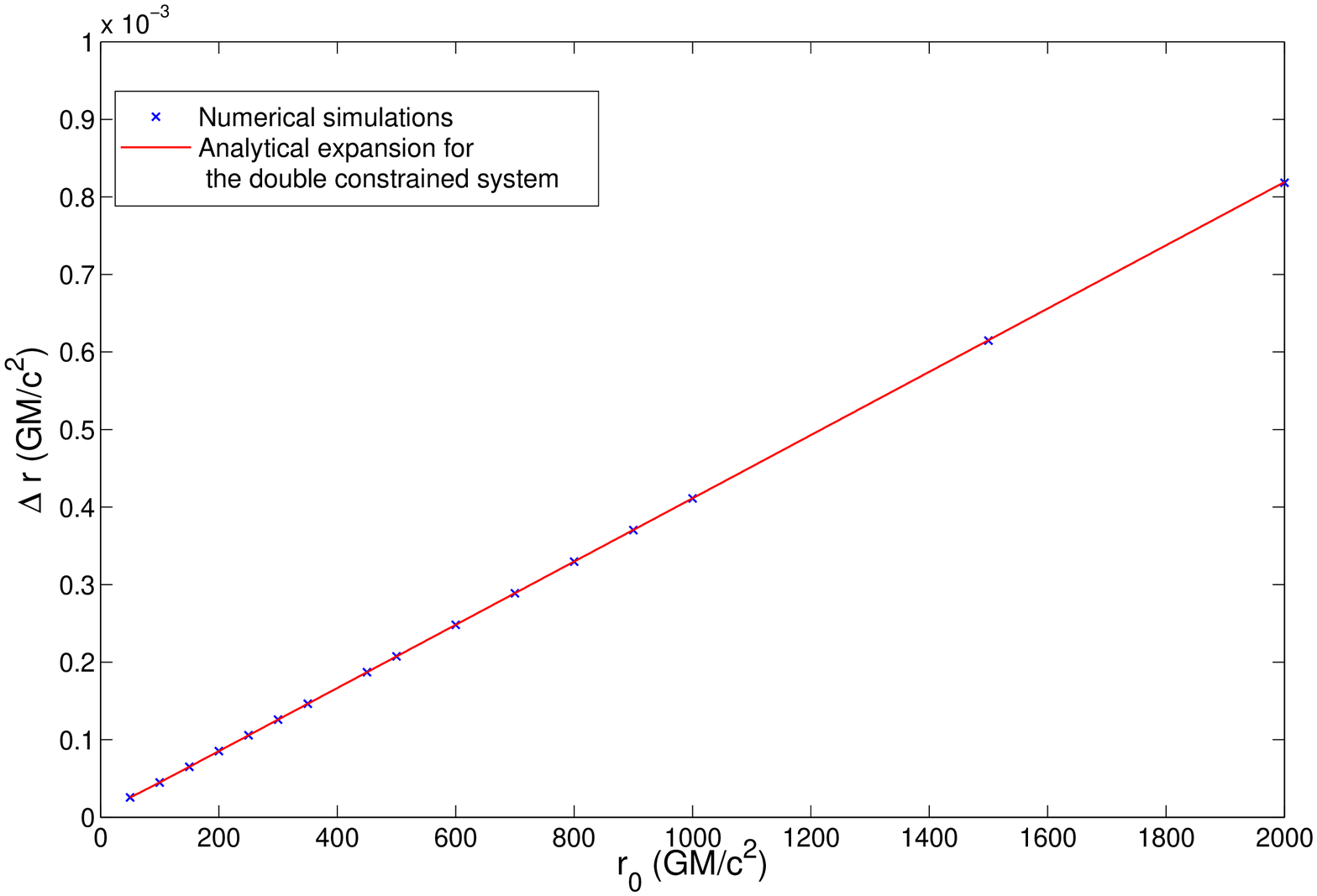}
\end{center}
\end{minipage}
\begin{minipage}{0.48\linewidth}
\begin{center}
		\includegraphics[width=1.0\linewidth]{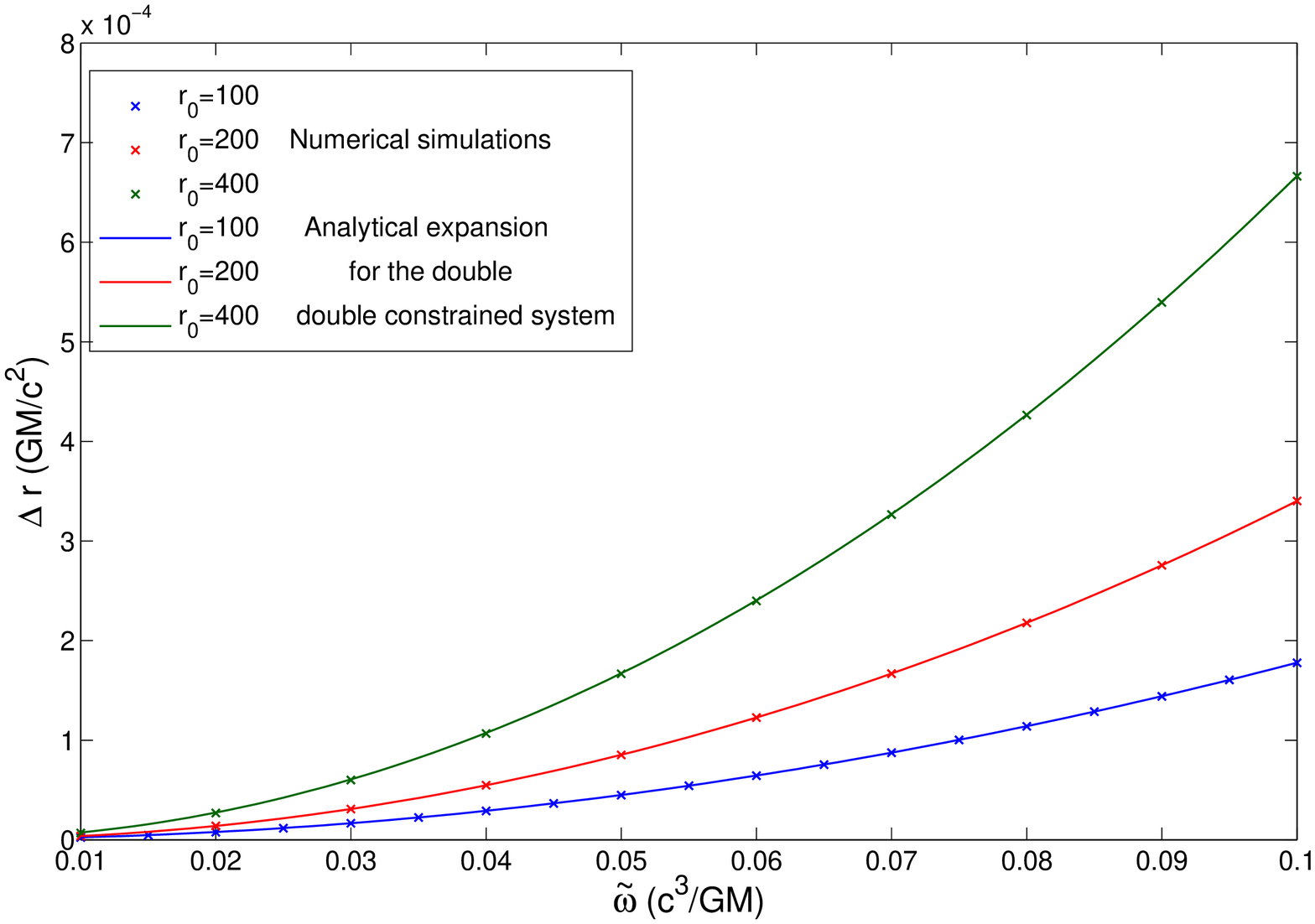}
\end{center}
\end{minipage}
	\label{figDrR0F}
	\caption{Comparison between the numerical simulations for the full system and the analytical expansion~(\ref{analExp}) for the double constrained system: (left)~$\Delta r$ as a function of $r_0$ for $\tilde{\omega}=0.05$, $\alpha=0$, $\delta_l=5 \times 10^{-3}$ and $l_0=5 \times 10^{-3}$; (right)~$\Delta r$ as a function of $\tilde{\omega}$ for different $r_0$ with $\alpha=0$, $\delta_l=5 \times 10^{-3}$ and $l_0=5 \times 10^{-3}$.}
\end{figure}

After a rather detailed analysis of the double constrained system, some results about the full system are presented. In this section the constraint $\theta = 0$ is relaxed. In other words the two-body system needs no longer be aligned radially, but can freely spin. Still, as initial conditions a radial alignment is chosen. With this choice of initial conditions the angle $\beta$ (see Fig.~\ref{figSys1}) will oscillate at a characteristic frequency. It is no longer possible to get analytical formulas as in the previous section, since in the equations of motion all variables are coupled and it is no longer possible to decouple them. Still, we can integrate numerically for rather small radii the full equations of motion.

In these simulations the effect seen in the double constrained system, namely the difference in the radius $\Delta r$, is found as well. This result is presented in Fig.~\ref{figDrR0F}, where the solid line is the analytical expression for the \emph{double constrained system}. As can be seen, the difference between the two systems remains unimportant. The same applies if $\Delta r$ is shown as a function of the frequency of the constraint~$\tilde \omega$.

This result is shown in Figure~\ref{figDrR0F} and again the analytical result from the double constrained system fits perfectly well the simulation points of the full system. Thus, as far as $\Delta r$ is concerned, the above results of the double constrained system describe this more complicated situation very well.

Still, on top of the shift in the radius the relative position of the system, described by the angle $\beta$, will oscillate around its equilibrium position $\beta = 0$. In the following, we are mainly interested in this new oscillation, which is characterized by its amplitude $\Delta \beta$ and its frequency $\omega_\beta$.

\begin{figure}[t]
\begin{minipage}{0.48\linewidth}
\begin{center}
		\includegraphics[width=1.0\linewidth]{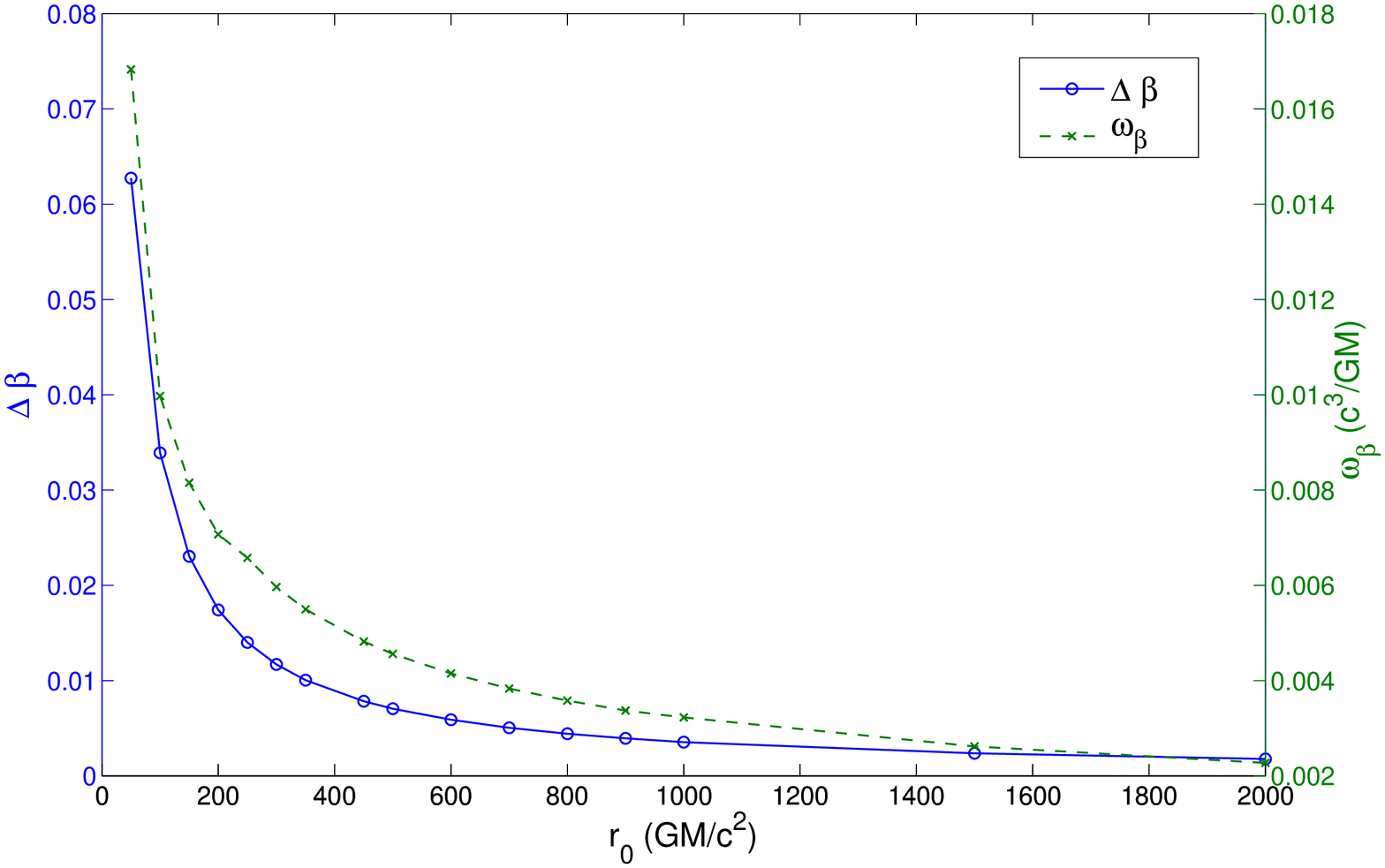}
\end{center}
\end{minipage}
\begin{minipage}{0.48\linewidth}
\begin{center}
		\includegraphics[width=1.0\linewidth]{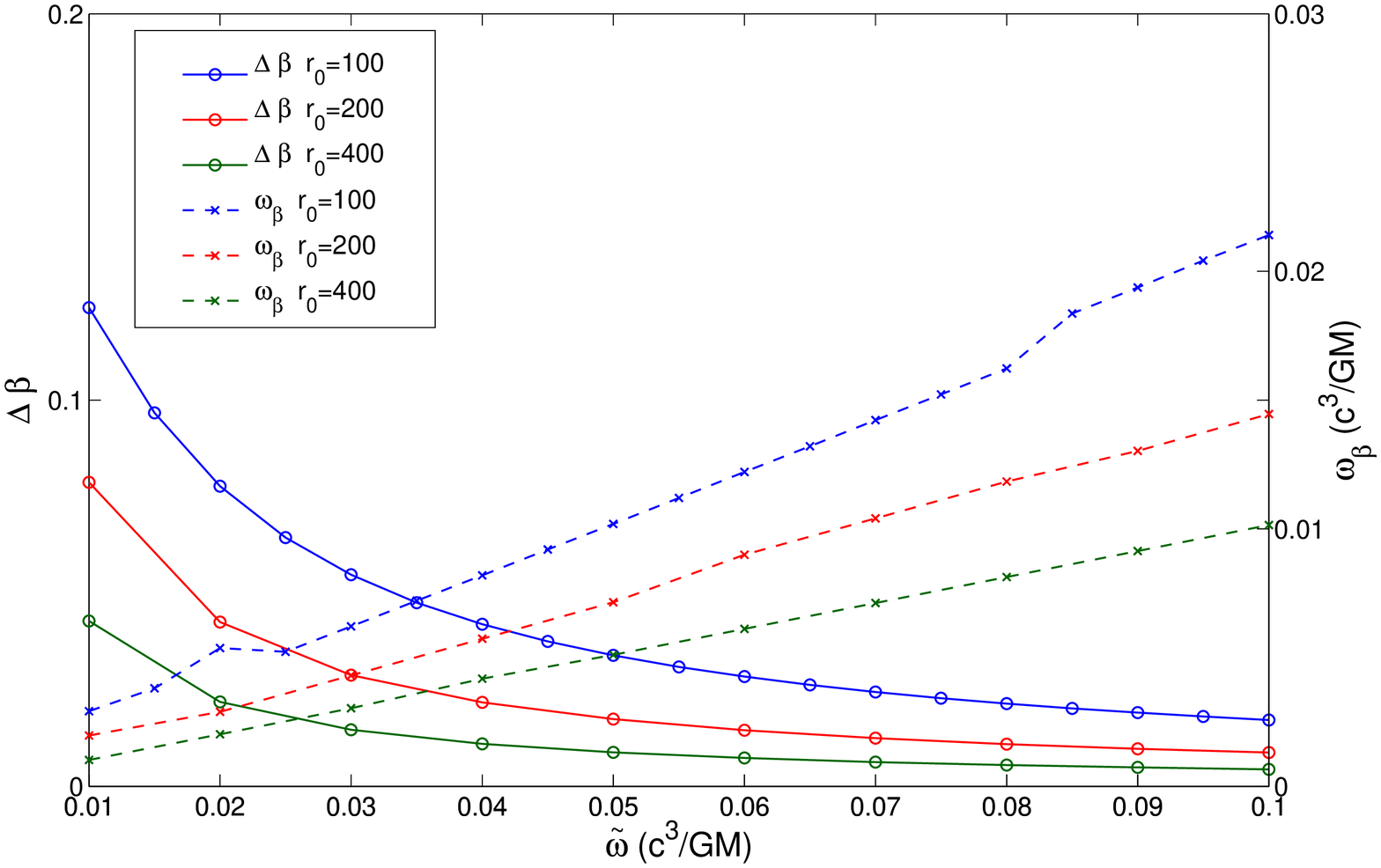}
\end{center}
\end{minipage}
	\label{figDbR0}
	\caption{Representation of $\Delta \beta$ and $\omega_\beta$ for the full system~: (left)~as a function of $r_0$ for $\tilde{\omega}=0.05$, $\alpha=0$, $\delta_l=5 \times 10^{-3}$ and $l_0=5 \times 10^{-3}$; (right)~as a function of $\tilde{\omega}$ for different $r_0$ with $\tilde{\omega}=0.05$, $\alpha=0$, $\delta_l=5 \times 10^{-3}$ and $l_0=5 \times 10^{-3}$.}
\end{figure}

Figure \ref{figDbR0} shows the behavior of $\Delta \beta$ and $\omega_\beta$ with the radius $r_0$. They are decreasing with the distance from the central body, as can be expected on general grounds since the gravity gradient decreases with increasing radius. As a consequence, we expect that at large radii the full system will behave like the double constrained system.

Finally, Figure \ref{figDbR0} shows the behavior of $\Delta r$, $\Delta \beta$ and $\omega_\beta$ as a function of the oscillation frequency $\tilde{\omega}$. It is seen that the amplitude of oscillation of the system around its center decreases with $\tilde{\omega}$ but the frequency of oscillation $\omega_\beta$ seems to increase linearly.

\section{Some comments on possible experiments}\label{secExp}

\begin{figure}[t]
	\centering
		\includegraphics[width=0.8\textwidth]{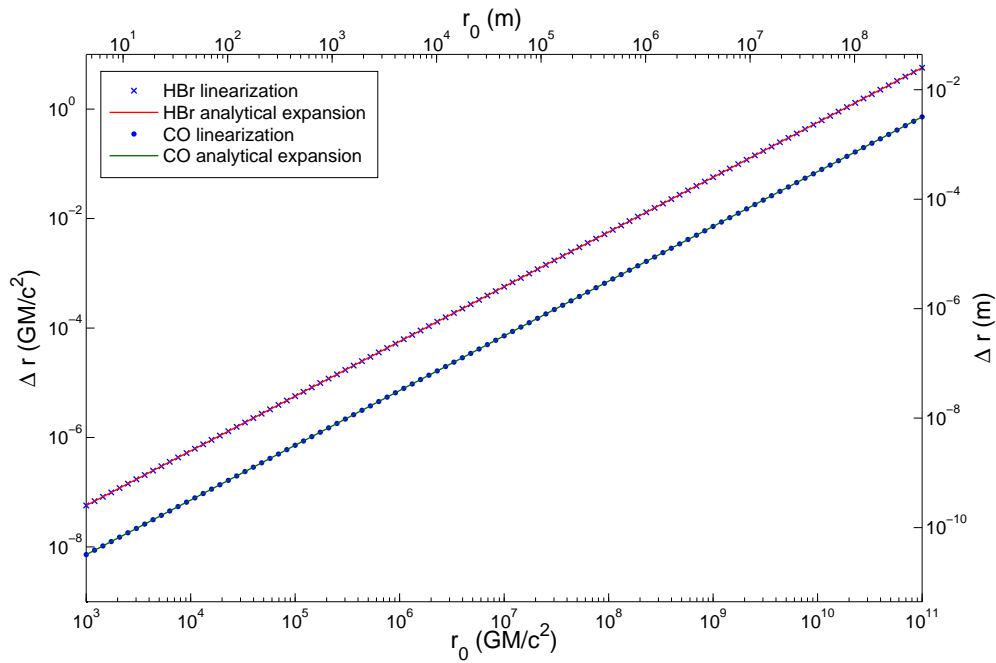}
	\caption{$\Delta r$ as a function of the initial radius $r_0$ for the double constrained system with values corresponding to the ground state molecular vibration of $HBr$ and $CO$ (see Tab \ref{tabMol}.) For the simulation of the linearized system \eqref{drLin} a central body whose mass is equal to the Earth's mass ($M=M_\oplus$) has been assumed, the analytical result is independent of $M$.}
	\label{figHighR0}
\end{figure}

Could the effect described in this paper be measured on-board a spacecraft orbiting the Earth? Since we mainly deal with a high velocity effect, the weak gravitational field does not form an obstacle. On the contrary, the maximum deviation from the reference motion grows linearly with the radius, which on the other hand is at the price of an increased integration time (half a revolution period). A promising system to measure the effect could be molecule vibrations; in the following we estimate the result from such a system for an idealized situation, where a single molecule is placed in orbit with its vibration direction exactly aligned radially. Using a semi-classical model, we can describe a molecule as a quantified anharmonic oscillator. With this simple model, we find the values of the oscillation parameters as presented in Table \ref{tabMol}.

\begin{table}[hb]
	\centering
			\begin{tabular}{cccccr}
		\hline
		\hline
		 & $l_0$ & $\tilde{\omega}$	& $\delta_l$ & $\delta_l^2\tilde{\omega}^2/c^2$ &  	  \\
		 &($GM/c^2$) & ($c^3/GM$)	& ($GM/c^2$)&  &  	  \\\hline
		 \multirow{2}{*}{$HBr$} & \multirow{2}{*}{$3.1 \times 10^{-8}$} & \multirow{2}{*}{1180} & $2.5 \times 10^{-9}$  & $8.7 \times 10^{-12}$ &\\
		        &              &       & $2.7 \times 10^{-8}$ & $1.0 \times 10^{-9}$ & (excited) \\
		 \multirow{2}{*}{$CO$} & \multirow{2}{*}{$2.5 \times 10^{-8}$} &   \multirow{2}{*}{ 954}& $1.1 \times 10^{-9}$  & $1.1 \times 10^{-12}$ &\\
		      &               &               & $1.9 \times 10^{-8}$ & $3.2 \times 10^{-10}$& (excited)\\ \hline \hline			
		\end{tabular}
	\caption{Value of the oscillation parameters for two different molecule vibrations. $M$ is the mass of the Earth.}
	\label{tabMol}
\end{table}

Notice that these values are given in geometrical units, with $M$ the mass of the Earth; therefore specific numbers are only valid in this environment. Nevertheless, the leading contribution in \eqref{analExp} only depends on the velocity $\delta_l \tilde \omega$, which is independent of $M$. Therefore the results presented here to a very high accuracy hold in a (weak) gravitational field different from the one of the Earth.

If the molecule is taken in its ground state, the result as depicted in Figure~\ref{figHighR0} is obtained, where both, the analytical expression as well as simulation points for the linearized system, are plotted. We can see that for Earth radius and low Earth orbits ($r_0\approx1.43 \ 10^9 GM/c^2$), the effect is of the order of 
\begin{equation}
\Delta r \sim 10^{-2} GM/c^2 \sim 5 \times 10^{-2} \mbox{mm.}
\end{equation}
The result does not look promising, but there exist different strategies to increase the effect. Firstly, one might try to work with excited states, which can increase the effect by about a factor of 100. Secondly, the experiment could be placed in a higher orbit. As an example, one can win a factor of 10 by placing the experiment in a geostationary orbit. Taking both strategies together, an effect of about 5~cm results. Of course, one might also try to place a spacecraft in an orbit around a central body different from the Earth. If the experiment is placed in an orbit around the sun with a radius equal to the one of the semimajor axis of the Earth's orbit, we get a displacement of about 10~m for the ground-state vibrations, and about 1~km for excited states.

As has been mentioned in Sect.~\ref{sec:analytical} besides the shift $\Delta r$ the perturbed trajectory is characterized by an advance of the periapsis of the ellipse. We mention that for molecular vibrations the typical values for this advance are of the order of $10^{-12}$~rad and thus not of experimental interest.


\section{Conclusions}
\label{conclusions}

In this paper we investigated the general relativistic effects of vibrations on a two-body system placed in an orbit around a central body. In our analysis the background spacetime was taken as Schwarzschild spacetime and the reference motion of the two-body system was assumed to be a circular orbit. Within a suitable expansion of the true motion around the reference motion, the equations of motion can be solved analytically, which led to our main result the relation~\eqref{analExp}. It was found that the vibrations deform the circular orbit to an ellipse, described by the maximal deviation of the reference motion and an advance of the periapsis. We stressed the similarities and differences between the experiment we described with the one by Shirokov~\cite{Shirokov:1973gr}.

In general relativity, the dominant contribution to the deformation is a high-velocity effect and the maximal deviation from the reference motion grows linearly with the radius $r_0$ thereof. In contrast to this, no high-velocity effects are present in the Newtonian theory and the dominant effect decays like $1/r_0$. Since the relativistic effects are at large radii orders of magnitude larger than the Newtonian one, it should be possible to measure this effect in the Solar system.

An estimate of the orders of magnitude has been presented for molecular vibrations. A promising strategy could be to place a spacecraft in a high orbit (geostationary or higher) and to use excited states as vibrations instead of the ground state. Of course, many questions considering possible experiments are open. Here we considered an idealized situation where a single molecule is placed in a circular orbit. Moreover, we did not take into account that the two atoms within the molecule can have a different mass. More realistic situations, possibly including a statistical analysis over a large number of vibrating molecules, should be analyzed. Also, different experimental setups could be studied, e.g.\ vibrations in crystals or nanoparticles. However, for most of these situations more complex models than presented here should be studied. Further topics could be the use of more complicated background spacetimes and different orbits as reference motions.


\paragraph*{Acknowledgments} The authors would like to thank D.~Izzo for important discussions on the topic. A.~Hees is research fellow from FRS-FNRS (Belgian Fund for Scientific Research) and he thanks FRS-FNRS for financial support for his thesis at ORB-UCL (Observatoire Royal de Belgique - Universit\'e Catholique de Louvain, Belgium).


\bibliographystyle{apsrev}
\bibliography{../../../literature}

\begin{thebibliography}{10}
\expandafter\ifx\csname natexlab\endcsname\relax\def\natexlab#1{#1}\fi
\expandafter\ifx\csname bibnamefont\endcsname\relax
  \def\bibnamefont#1{#1}\fi
\expandafter\ifx\csname bibfnamefont\endcsname\relax
  \def\bibfnamefont#1{#1}\fi
\expandafter\ifx\csname citenamefont\endcsname\relax
  \def\citenamefont#1{#1}\fi
\expandafter\ifx\csname url\endcsname\relax
  \def\url#1{\texttt{#1}}\fi
\expandafter\ifx\csname urlprefix\endcsname\relax\def\urlprefix{URL }\fi
\providecommand{\bibinfo}[2]{#2}
\providecommand{\eprint}[2][]{\url{#2}}

\bibitem[{\citenamefont{Shirokov}(1973)}]{Shirokov:1973gr}
\bibinfo{author}{\bibfnamefont{M.}~\bibnamefont{Shirokov}},
  \bibinfo{journal}{Gen. Rel. Grav.} \textbf{\bibinfo{volume}{4}},
  \bibinfo{pages}{131} (\bibinfo{year}{1973}).

\bibitem[{\citenamefont{Martinez-Sanchez and Gavit}(1987)}]{Martinez:1987om}
\bibinfo{author}{\bibfnamefont{M.}~\bibnamefont{Martinez-Sanchez}}
  \bibnamefont{and} \bibinfo{author}{\bibfnamefont{S.}~\bibnamefont{Gavit}},
  \bibinfo{journal}{J. Guid. Control Dyn.} \textbf{\bibinfo{volume}{10}},
  \bibinfo{pages}{233} (\bibinfo{year}{1987}).

\bibitem[{\citenamefont{Landis and Hrach}(1991)}]{Landis:1991sr}
\bibinfo{author}{\bibfnamefont{G.}~\bibnamefont{Landis}} \bibnamefont{and}
  \bibinfo{author}{\bibfnamefont{F.}~\bibnamefont{Hrach}}, \bibinfo{journal}{J.
  Guid. Control Dyn.} \textbf{\bibinfo{volume}{14}}, \bibinfo{pages}{214}
  (\bibinfo{year}{1991}).

\bibitem[{\citenamefont{Landis}(1992)}]{Landis:1992ro}
\bibinfo{author}{\bibfnamefont{G.}~\bibnamefont{Landis}},
  \bibinfo{journal}{Acta Astronautica} \textbf{\bibinfo{volume}{26}},
  \bibinfo{pages}{307} (\bibinfo{year}{1992}).

\bibitem[{\citenamefont{Wisdom}(2003)}]{Wisdom:2003aa}
\bibinfo{author}{\bibfnamefont{J.}~\bibnamefont{Wisdom}},
  \bibinfo{journal}{Science} \textbf{\bibinfo{volume}{299}},
  \bibinfo{pages}{2865} (\bibinfo{year}{2003}).

\bibitem[{\citenamefont{Gueron et~al.}(2006)\citenamefont{Gueron, Maia, and
  Matsas}}]{Gueron:2005ye}
\bibinfo{author}{\bibfnamefont{E.}~\bibnamefont{Gueron}},
  \bibinfo{author}{\bibfnamefont{C.~A.~S.} \bibnamefont{Maia}},
  \bibnamefont{and} \bibinfo{author}{\bibfnamefont{G.~E.~A.}
  \bibnamefont{Matsas}}, \bibinfo{journal}{Phys. Rev.}
  \textbf{\bibinfo{volume}{D73}}, \bibinfo{pages}{024020}
  (\bibinfo{year}{2006}), \eprint{gr-qc/0510054}.

\bibitem[{\citenamefont{Gueron and Mosna}(2007)}]{Gueron:2006fq}
\bibinfo{author}{\bibfnamefont{E.}~\bibnamefont{Gueron}} \bibnamefont{and}
  \bibinfo{author}{\bibfnamefont{R.~A.} \bibnamefont{Mosna}},
  \bibinfo{journal}{Phys. Rev.} \textbf{\bibinfo{volume}{D75}},
  \bibinfo{pages}{081501} (\bibinfo{year}{2007}), \eprint{gr-qc/0612.131}.

\bibitem[{\citenamefont{Manasse and Misner}(1962)}]{Misner:1962}
\bibinfo{author}{\bibfnamefont{F.~K.} \bibnamefont{Manasse}} \bibnamefont{and}
  \bibinfo{author}{\bibfnamefont{C.~W.} \bibnamefont{Misner}},
  \bibinfo{journal}{Journal of Mathematical Physics}
  \textbf{\bibinfo{volume}{4}}, \bibinfo{pages}{735} (\bibinfo{year}{1962}).

\bibitem[{\citenamefont{Melkumova and Khlebnikov}(1990)}]{Melkumova:1990rp}
\bibinfo{author}{\bibfnamefont{E.}~\bibnamefont{Melkumova}} \bibnamefont{and}
  \bibinfo{author}{\bibfnamefont{V.}~\bibnamefont{Khlebnikov}},
  \bibinfo{journal}{Russian Physics Journal} \textbf{\bibinfo{volume}{33}},
  \bibinfo{pages}{349} (\bibinfo{year}{1990}).

\bibitem[{\citenamefont{Vladimirov and Rodichev}(1981)}]{Vladimirov:1981rp}
\bibinfo{author}{\bibfnamefont{Y.}~\bibnamefont{Vladimirov}} \bibnamefont{and}
  \bibinfo{author}{\bibfnamefont{S.}~\bibnamefont{Rodichev}},
  \bibinfo{journal}{Russian Physics Journal} \textbf{\bibinfo{volume}{24}},
  \bibinfo{pages}{954} (\bibinfo{year}{1981}).

\end{thebibliography}

\end{document}